\documentclass{article}
\usepackage{amsmath}
\usepackage{graphicx}
\usepackage{graphics}

\input{tcilatex}
\begin{document}

\title{Lepton Polarization\ Asymmetry in $B\rightarrow l^{+}l^{-}$ decays in
R-parity violating Minimal Supersymmetric Standard Model}
\author{\ \ \ $Azeem\ Mir\footnotemark $,$\ Farida\ Tahir\footnotemark $, $%
Kamaluddin\ Ahmed\footnotemark \ \ \ $ \\
(\textit{Physics Department, COMSATS Institute of Information Technology,
Islamabad.})}
\date{}
\maketitle

\begin{abstract}
\textit{We study the implication of R-parity violating (}$\NEG{R}_{p}$%
\textit{) Minimal Supersymmetric Standard Model (MSSM) model in lepton
polarization asymmetry (}$A_{LP}$\textit{) in}$\ B\rightarrow l^{+}l^{-}$%
\textit{\ decays}$\ $\textit{. The analysis show that} \textit{the} $\mathit{%
A}_{LP}$\textit{\ is significant in a certain phenomenological parametric
region of Yukawa couplings. We have also placed indirect bounds on }$%
(\lambda ^{\prime \ast }\lambda )$ couplings as obtained from $B\rightarrow
\tau ^{+}\tau ^{-}.$
\end{abstract}

\footnotetext[1]{%
neutrino\_79@hotmail.com}\footnotetext[2]{%
farida\_tahir@comsats.edu.pk}\footnotetext[3]{%
kamal@comsats.edu.pk, kahmed\_pk@yahoo.com} Rare leptonic and semileptonic B
decays are a sensitive probe of physics beyond the standard model (SM).
These decays are suppressed in SM and proceed through higher order diagrams.
One can also study these decays in various extensions of SM like Two Higgs
Doublet Model (2HDM) and $\NEG{R}_{p}\ $MSSM \cite{CH}$.$Rare$\ B$ decays
are being searched at $B$ factories $(Belle$ and$\ BaBar)$ \cite{BFACT} and
also at Tevatron ($CDF$ and $DO$) \cite{TEV}. MSSM \ provides a framework
for these decays at tree level. These decays provide valuable information
through various phenomenological observables. $A_{LP}$ is one such
observable. It is zero in SM and is non zero only in the presence of scalar
and pseudoscalar interactions. Therefore its observation would provide a
direct evidence of existence of such interactions. Some authors have
explored the possibility of $A_{LP}$ in Higgs-Doublet Model \cite{LT}. We
explore this possibility in the frame work of $\NEG{R}_{p}$ MSSM. Here,\
these interactions arise naturally as sneutrino and squark exchange terms.
The $\NEG{R}_{p}$ superpotential is \cite{H1}.%
\begin{equation}
\ W_{\text{\textsl{$\;\NEG{R}$}}_{\text{\textbf{p}}}}=\frac{1}{2}\lambda
_{ijk}L_{i}L_{j}E_{k}^{c}+\lambda _{ijk}^{^{\prime }}L_{i}Q_{j}D_{k}^{c}+%
\frac{1}{2}\lambda _{ijk}^{^{\prime \prime
}}U_{i}^{^{C}}D_{j}^{^{C}}D_{k}^{^{C}}
\end{equation}%
where $i,\,j,\,k$\ are generation indices, $L_{i}$\ and $Q_{i}$\ are the
lepton and quark left-handed $SU(2)_{L}$\ doublets and $E^{c}$, $D^{c}$\ are
the charge conjugates of the right-handed leptons and quark singlets,
respectively. $\lambda _{ijk}$\ , $\lambda _{ijk}^{\prime }$\ \ and \ $%
\lambda _{ijk}^{\prime \prime }$ are Yukawa couplings. Note that the term
proportional to $\lambda _{ijk}$\ is antisymmetric in first two indices $%
[i,\,j]$\ and $\lambda _{ijk}^{\prime \prime }$ is antisymmetric in\ last$\ $%
two indices $[j,k]$, implying $9(\lambda _{ijk})+27(\lambda _{ijk}^{^{\prime
}})+9\left( \lambda _{ijk}^{^{\prime \prime }}\right) =45$\ independent
coupling constants among which 36 are related to the lepton flavor violation
(9 from $LLE^{c}$\ and 27 from $LQD^{c}$).\ The matrix element of $%
(B\rightarrow l_{\beta }^{+}l_{\beta }^{-})$ is given as%
\begin{equation}
M=\frac{G_{F}\alpha V_{tb}V_{tq}^{\ast }}{\sqrt{2}}(C_{PP}\ (\overline{q}%
\gamma ^{5}b)(\overline{l_{\beta }}\gamma ^{5}l_{\beta })+C_{PS}\ (\overline{%
q}\gamma ^{5}b)(\overline{l_{\beta }}l_{\beta })+C_{AA}\ (\overline{q}\gamma
^{\mu }\gamma ^{5}b)(\overline{l_{\beta }}\gamma ^{\mu }\gamma ^{5}l_{\beta
}))
\end{equation}

where $q=d,s\ ;\ \beta =e,\mu ,\tau $ and $C_{PP},\ C_{PS},\ C_{AA}$ are
pseudoscalar, scalar and axial form factors as given in \cite{GR}

We reproduce here the following results on branching ratio and $A_{LP}$
defined in \cite{LT} after using PCAC.%
\begin{eqnarray}
Br\left[ B_{q}\rightarrow l_{\beta }^{+}l_{\beta }^{-}\right]  &=&\frac{%
G_{F}^{2}\alpha ^{2}}{16\pi }\left\vert V_{tb}V_{tq}^{\ast }\right\vert
^{2}(f_{B_{q}})^{2}m_{B_{q}}\tau _{B}\sqrt{(1-(\frac{2m_{l_{\beta }}}{%
m_{B_{q}}})^{2})}  \notag \\
\lbrack  &\mid &2m_{l_{\beta }}C_{AA}-\frac{(m_{B_{q}})^{2}}{m_{b}+m_{q}}%
C_{PP}\mid ^{2}+(1-(\frac{2m_{l_{\beta }}}{m_{B_{q}}})^{2})\left\vert \frac{%
(m_{B_{q}})^{2}}{m_{b}+m_{q}}C_{PS}\right\vert ^{2}]
\end{eqnarray}

The expression for longitudinal polarization\ asymmetry ($A_{LP}$) is given
by%
\begin{equation}
A_{LP}=\frac{2\frac{(m_{B_{q}})}{m_{b}+m_{q}}\ \sqrt{1-(\frac{2m_{l_{\beta }}%
}{m_{B_{q}}})^{2}}\func{Re}[C_{PS}(2\frac{m_{l}}{m_{B_{q}}}C_{AA}-\frac{%
(m_{B_{q}})}{m_{b}+m_{q}}C_{PP})^{\ast }]}{\mid \frac{2m_{l_{\beta }}}{%
m_{B_{q}}}C_{AA}-\frac{(m_{B_{q}})}{m_{b}+m_{q}}C_{PP}\mid ^{2}+(1-(\frac{%
2m_{l_{\beta }}}{m_{B_{q}}})^{2})\left\vert \frac{(m_{B_{q}})}{m_{b}+m_{q}}%
C_{PS}\right\vert ^{2}}
\end{equation}

One re-expresses $A_{LP}$ (eq(4)) in terms of branching fraction (eq(3)) and 
$C_{PS}$, taking the form factors to be real for practical purposes.%
\begin{equation}
A_{LP}=\frac{2\sqrt{h}}{Br[B_{q}\rightarrow l_{\beta }^{+}l_{\beta }^{-}]}%
\func{Re}[C_{PS}\sqrt{Br(B_{q}\rightarrow l_{\beta }^{+}l_{\beta
}^{-})-h\left\vert C_{PS}\right\vert ^{2}}]
\end{equation}%
where the constant $h$ is defined as%
\begin{equation}
h=\frac{G_{F}^{2}\alpha ^{2}}{16\pi }\left\vert V_{tb}V_{tq}^{\ast
}\right\vert ^{2}(f_{B_{q}})^{2}m_{B_{q}}\tau _{B_{q}}(1-(\frac{2m_{l_{\beta
}}}{m_{B_{q}}})^{2})(\frac{(m_{B_{q}})^{2}}{m_{b}+m_{q}})^{2}
\end{equation}%
Eq.(5) can be re-arranged%
\begin{equation}
Br[B_{q}\rightarrow l_{\beta }^{+}l_{\beta }^{-}]=\frac{2h}{1\pm \sqrt{%
1-(A_{LP})^{2}}}\left\vert C_{PS}\right\vert ^{2}
\end{equation}%
We discuss the possibility of non-zero $A_{LP\text{ }}$in $(B_{q}\rightarrow
l_{\beta }^{+}l_{\beta }^{-})$\ in $\NEG{R}_{p}\ $framework. We show that it
arises naturally by sneutrino exchange terms.\ 

In $\NEG{R}_{p}$ MSSM the relevant effective Lagrangian is given by \cite%
{H1,FAR}

\begin{equation}
L_{\QTR{sl}{\NEG{R}\;}_{\text{\textbf{p}}}}^{eff}\left( \ \overline{b}%
q\longrightarrow l_{\beta }+\overline{l}_{\beta }\right) =\frac{4G_{F}\alpha
V_{tb}V_{tq}^{\ast }}{\sqrt{2}}\left[ 
\begin{array}{c}
A_{\beta \beta }^{bq}\left( \overline{l_{\beta }}\gamma ^{\mu }P_{L}l_{\beta
}\right) \left( \overline{q}\gamma _{\mu }P_{R}b\right) \\ 
-B_{\beta \beta }^{bq}\left( \overline{l_{\beta }}P_{R}l_{\beta }\right)
\left( \overline{q}P_{L}b\right) \\ 
-C_{\beta \beta }^{bq}\left( \overline{l_{\beta }}P_{L}l_{\beta }\right)
\left( \overline{q}P_{R}b\right)%
\end{array}%
\right]
\end{equation}%
The first term in eq. (8) comes from the up squark exchange (where $q$ and $%
b $\ are down type quarks). The dimensionless coupling constants $A_{\beta
\beta }^{bq},$ $B_{\beta \beta }^{bq}\ $and $C_{\beta \beta }^{bq}\ $depend
on the species of charged leptons and are given by

\begin{equation}
A_{\beta \beta }^{bq}=\frac{\sqrt{2}}{4G_{F}\alpha }\underset{m,n,i=1}{%
\overset{3}{\sum }}\frac{V_{ni}^{\dagger }V_{im}}{V_{tb}V_{tq}^{\ast }}\frac{%
\lambda _{\beta n3}^{\prime }\lambda _{\beta mq}^{\prime \ast }}{2m_{%
\widetilde{u_{i}^{c}}}^{2}}
\end{equation}%
\begin{equation}
B_{\beta \beta }^{bq}=\frac{\sqrt{2}}{4G_{F}\alpha }\underset{i=1}{\overset{3%
}{\sum }}\frac{1}{V_{tb}V_{tq}^{\ast }}\frac{2\lambda _{i\beta \beta }^{\ast
}\lambda _{iq3}^{\prime }}{m_{\widetilde{\nu }_{iL}}^{2}}
\end{equation}%
\begin{equation}
C_{\beta \beta }^{bq}=\frac{\sqrt{2}}{4G_{F}\alpha }\underset{i=1}{\overset{3%
}{\sum }}\frac{1}{V_{tb}V_{tq}^{\ast }}\frac{2\lambda _{i\beta \beta
}\lambda _{i3q}^{\prime \ast }}{m_{\widetilde{\nu _{iL}}}^{2}}
\end{equation}%
Form factors are obtained by comparing Eq. (2) by Eq. (8).

\begin{equation}
C_{AA}=A_{\beta \beta }^{bq}
\end{equation}%
\begin{equation}
C_{PS}=B_{\beta \beta }^{bq}+C_{\beta \beta }^{bq}
\end{equation}%
\begin{equation}
C_{PP}=B_{\beta \beta }^{bq}-C_{\beta \beta }^{bq}
\end{equation}

\ \ For $\beta =e,\mu ;\ \frac{m_{l_{\beta }}}{m_{B_{q}}}\simeq 0.$eq.(4).
reduces to%
\begin{equation}
A_{LP}=\frac{-2\func{Re}[C_{PS}C_{PP}^{\ast }]}{\mid C_{PP}\mid
^{2}+\left\vert C_{PS}\right\vert ^{2}}
\end{equation}%
Substituting the expressions of form factors from eq.(13,14) into eq.(15),
we get after simplifying%
\begin{equation}
A_{LP}=\frac{\left\vert \frac{\lambda _{i\beta \beta }\lambda _{i3q}^{\prime
\ast }}{m_{\widetilde{\nu }_{Li}}^{2}}\right\vert ^{2}-\left\vert \frac{%
\lambda _{i\beta \beta }^{\ast }\lambda _{iq3}^{\prime }}{m_{\widetilde{\nu }%
_{Li}}^{2}}\right\vert ^{2}}{\left\vert \frac{\lambda _{i\beta \beta
}\lambda _{i3q}^{\prime \ast }}{m_{\widetilde{\nu }_{Li}}^{2}}\right\vert
^{2}+\left\vert \frac{\lambda _{i\beta \beta }^{\ast }\lambda _{iq3}^{\prime
}}{m_{\widetilde{\nu }_{Li}}^{2}}\right\vert ^{2}}
\end{equation}%
Eq.(18) may be used to constrain the parameter space $(\lambda
_{i3q}^{\prime \ast }\lambda _{i\beta \beta },\lambda _{iq3}^{\prime
}\lambda _{i\beta \beta }^{\ast })$ for observed values of $A_{LP}$. It also
shows that$\ A_{LP}$ is observable for$\ B\rightarrow l_{\beta }^{+}l_{\beta
}^{-}\ (\beta =e,\mu )$ if size of $\lambda _{iq3}^{\prime }\lambda _{i\beta
\beta }^{\ast }$ is different from $\lambda _{i3q}^{\prime \ast }\lambda
_{i\beta \beta }$ (see Fig.1) .

We also get after substituting form factors from eq.(12-14) into eq. (5).

\begin{equation}
A_{LP}=\frac{2\sqrt{h^{\prime }}}{Br[B_{q}\rightarrow l_{\beta }^{+}l_{\beta
}^{-}]}\func{Re}[(\frac{\lambda _{i\beta \beta }^{\ast }\lambda
_{iq3}^{\prime }+\lambda _{i\beta \beta }\lambda _{i3q}^{\prime \ast }}{%
2(m_{v_{iL}})^{2}})\sqrt{Br(B_{q}\rightarrow l_{\beta }^{+}l_{\beta }^{-})-%
\frac{h^{\prime }}{4}\left\vert \frac{\lambda _{i\beta \beta }^{\ast
}\lambda _{iq3}^{\prime }+\lambda _{i\beta \beta }\lambda _{i3q}^{\prime
\ast }}{(m_{v_{iL}})^{2}}\right\vert ^{2}}]
\end{equation}

\begin{equation}
Br[B_{q}\rightarrow l_{\beta }^{+}l_{\beta }^{-}]=\frac{2h^{\prime }}{1\pm 
\sqrt{1-(A_{LP})^{2}}}\left\vert \frac{\lambda _{i\beta \beta }^{\ast
}\lambda _{iqb}^{\prime }+\lambda _{i\beta \beta }\lambda _{ibq}^{\prime
\ast }}{2(m_{v_{iL}})^{2}}\right\vert ^{2}
\end{equation}

where%
\begin{equation}
h^{\prime }=\frac{1}{16\pi }(f_{B_{q}})^{2}m_{B_{q}}\tau _{B_{q}}(1-(\frac{%
2m_{l_{\beta }}}{m_{B_{q}}})^{2})(\frac{(m_{B_{q}})^{2}}{m_{b}+m_{q}})^{2}
\end{equation}

Branching fraction for $(B_{q}\rightarrow l_{\beta }^{+}l_{\beta }^{-};\
\beta =e,\mu )$ are calculated \cite{LT}.%
\begin{equation}
Br(B_{s}^{0}\rightarrow l_{\beta }^{+}l_{\beta }^{-})=\left\{ 
\begin{array}{c}
8.9\times 10^{-14}(\frac{f_{B}}{245\ MeV})^{2},\beta =e \\ 
4.0\times 10^{-9}(\frac{f_{B}}{245\ MeV})^{2},\beta =\mu \\ 
8.3\times 10^{-7}(\frac{f_{B}}{245\ MeV})^{2},\beta =\tau%
\end{array}%
\right.
\end{equation}

\begin{center}
\begin{equation}
Br(B_{d}^{0}\rightarrow l_{\beta }^{+}l_{\beta }^{-})=\left\{ 
\begin{array}{c}
3.4\times 10^{-15}(\frac{f_{B}}{210\ MeV})^{2},\beta =e \\ 
1.5\times 10^{-9}(\frac{f_{B}}{210\ MeV})^{2},\beta =\mu \\ 
3.2\times 10^{-8}(\frac{f_{B}}{210\ MeV})^{2},\beta =\tau%
\end{array}%
\right.
\end{equation}
\end{center}

Experimental bounds on branching fraction are given \cite{pdg}

\begin{center}
$\ \ \ \ \ \ \ \ \ \ \ \ Br(B_{s}^{0}\rightarrow l_{\beta }^{+}l_{\beta
}^{-})<\left\{ 
\begin{array}{c}
5.4\times 10^{-5},\beta =e \\ 
1.5\times 10^{-7},\beta =\mu%
\end{array}%
\right. $\ \ \ \ \ \ \ \ \ \ \ \ \ 

\ $Br(B_{d}^{0}\rightarrow l^{+}l^{-})<\left\{ 
\begin{array}{c}
6.1\times 10^{-8},l=e \\ 
3.9\times 10^{-8},l=\mu%
\end{array}%
\right. $\ \ 
\end{center}

Indirect bounds on branching fractions from tau decays are given \cite{GI}

\begin{center}
\ $Br(B_{d}^{0}\rightarrow \tau ^{+}\tau ^{-})<6\times 10^{-6}\medskip $

\ $Br(B_{s}^{0}\rightarrow \tau ^{+}\tau ^{-})<2\times 10^{-5}$
\end{center}

We have studied the relation between branching fraction and Yukawa couplings 
$(\lambda ^{\ast }\lambda ^{\prime })$ in Figs.(2,3,4). We have used data
from \ \cite{pdg,SH}. The graphs show a constrained region where branching
fraction varies between upper experimental and lower theoretical limit for a
fixed value of $A_{LP}$. The plots show that for $(B_{q}\rightarrow l_{\beta
}^{+}l_{\beta }^{-};\beta =\mu ,\tau ;\ q=d,s)\ $Yukawa couplings have\ a
very tight parameter space as compared to $(B_{q}\rightarrow e^{+}e^{-})$.$\
\ $We have used\ bounds on Yukawa couplings \cite{HK} for our analysis

\begin{center}
$\left\vert \frac{\lambda _{232}^{\prime }\lambda _{211}}{(m_{\grave{v}%
_{2L}}/100GeV)^{2}}\right\vert <2.3\times 10^{-4};\ \left\vert \frac{\lambda
_{132}^{\prime }\lambda _{122}}{(m_{\grave{v}_{1L}}/100GeV)^{2}}\right\vert
<1.2\times 10^{-5}\medskip $

$\left\vert \frac{\lambda _{231}^{\prime }\lambda _{211}}{(m_{\grave{v}%
_{2L}}/100GeV)^{2}}\right\vert <4.1\times 10^{-5};\ \left\vert \frac{\lambda
_{131}^{\prime }\lambda _{122}}{(m_{\grave{v}_{1L}}/100GeV)^{2}}\right\vert
<6.2\times 10^{-6}$
\end{center}

Indirect bounds on Yukawa couplings $(\lambda ^{\ast }\lambda ^{\prime })$
can be obtained for $(B_{q}\rightarrow \tau ^{+}\tau ^{-})$ from\ Fig. (4)
assuming single coupling dominance.

\begin{center}
$\left\vert \frac{\lambda _{i31}^{^{\prime \ast }}\lambda _{i33}}{(m_{\grave{%
v}_{iL}}/100GeV)^{2}}\right\vert <8\times 10^{-5};\ \left\vert \frac{\lambda
_{i32}^{^{\prime \ast }}\lambda _{i33}}{(m_{\grave{v}_{iL}}/100GeV)^{2}}%
\right\vert <1.2\times 10^{-4};\ i=1,2$
\end{center}

Summarizing we have studied contribution of $\NEG{R}_{p}$ MSSM to $A_{LP}$.
Our analysis shows that $A_{LP}$ receives main contribution from
sneutrino-exchange term involving Yukawa couplings $(\lambda ^{\prime
}\lambda )$. The squark exchange term involving Yukawa couplings $(\lambda
^{\prime }\lambda ^{\prime })$ is suppressed by the lepton mass. An
experimental observation of $A_{LP}$ and its numerical value may help us to
constrain the parameter space of Yukawa couplings $(\lambda ^{\prime
}\lambda )$. We have also estimated indirect bounds on couplings $(\lambda
_{i31}^{^{\prime \ast }}\lambda _{i33},\lambda _{i32}^{^{\prime \ast
}}\lambda _{i33}).$Our predictions can be matched with future experimental
searches of rare $B$ decays at\ Tevatron,\ $B$ factories and at CERN ($LHC$
collider). The comparison may indirectly verify $\NEG{R}_{p}$ prediction
based on dominant sneutrino exchange process.

\subsection{\textbf{Acknowledgement}}

Azeem Mir is indebted to the Indigenous Ph.D. Fellowship Program of Higher
Education Commission of Pakistan.

\end{document}